\documentclass[twocolumn]{autart} 
\usepackage{graphicx}
\usepackage{amssymb}
\usepackage{amsmath}
\usepackage{cite}
\usepackage{color}

\newtheorem{secthm}{Theorem}[section]

\newtheorem{secdefn}[secthm]{Definition}
\newtheorem{secrem}[secthm]{Remark}
\def\red{\hfill $\lhd$}

\newcommand{\cC} { {\mathcal C}}
\newcommand{\cO} { {\mathcal O}}
\newcommand{\cR} { {\mathcal R}}
\newcommand{\bR} { {\mathbb R}}

\begin{document}
\allowdisplaybreaks[4]
\begin{frontmatter}
\title{Empirical Differential Gramians for\\ Nonlinear Model Reduction}

\thanks[footnoteinfo]{A preliminary version of this paper is presented at the 20th IFAC World Congress, July 2017. Corresponding author Y.~Kawano. Tel. +81 82 424 7582. 
Fax +81 82 422 7193.}

\author[JP]{Yu Kawano}\ead{ykawano@hiroshima-u.ac.jp},    % Add the 
\author[NL]{Jacquelien M.A. Scherpen}\ead{j.m.a.scherpen@rug.nl}

\address[JP]{Graduate School of Engineering, Hiroshima University, Kagamiyama 1-4-1, Higashi-Hiroshima 739-8527, Japan}

\address[NL]{Jan C. Willems Center for Systems and Control, Engineering and Technology institute Groningen, Faculty of Science and Engineering, University of Groningen, Nijenborgh 4, 9747 AG Groningen, the~Netherlands}


\begin{abstract}
In this paper, we present an empirical balanced truncation method for nonlinear systems with linear time-invariant input vector field components. First, we define differential reachability and observability Gramians. They are matrix valued functions of the state trajectory (i.e. the initial state and input trajectory) of the original nonlinear system, and it is difficult to find them as functions of the initial state and input. The main result of this paper is to show that for a fixed state trajectory, it is possible to compute the values of these Gramians by using impulse and initial state responses of the variational system. Therefore, balanced truncation is doable along the fixed state trajectory without solving nonlinear partial differential equations, differently from conventional nonlinear balancing methods. We further develop an approximation method, which only requires trajectories of the original nonlinear systems. Our methods are demonstrated by an RL network along a trajectory.
\end{abstract}
\end{frontmatter}

\section{Introduction}
Along with the development of new technologies, control systems are becoming more complex and large-scale. To capture systems' components which are essential for controller design and analysis, model order reduction techniques have been established, see e.g.~\cite{Antoulas:05}. In systems and control, typical methods are balanced truncation and moment matching~\cite{Antoulas:05,ZDG:96}, and both of them have been extended to nonlinear systems~\cite{Scherpen:93,FS:05,KS:17,BVSN:14,Astolfi:10,IA:16}. In contrast to successive theoretical developments, nonlinear model reduction methods still have computational challenges, since they require solutions to nonlinear partial differential equations (PDEs). There are few papers tackling this challenging problem such as~\cite{FT:08,SA:14,NK:00,SA:17,KBS:19}. As a data driven model order reduction method, proper orthogonal decomposition (POD)~\cite{HLB:12,Antoulas:05} is often used in practice. However, POD is mainly proposed for non-control systems.

For linear time-invariant (LTI) systems, POD and balancing are connected based on the fact that the controllability and observability Gramians can be computed by using impulse and initial state responses, respectively. That is, balanced truncation of LTI systems can be performed by using empirical data. Applying linear empirical methods to nonlinear systems have attracted various research interests, see e.g.,~\cite{LMG:02,HE:02,MR:04,Himpe:18,HE:02-2,WP:02}. Such methods are exploited in order to reduce the computational complexity of nonlinear controller design such as model predictive control~\cite{HKE:02,CSB:16}. 

However, these empirical methods have been proposed only around a steady-state because the aforementioned nonlinear balancing method gives the same reduced order model as the linear balancing method at a steady-state. For analysis and control of nonlinear systems, a steady-state is not always important. For instance, in a trajectory tracking control problem, a reduced order model around the trajectory could be useful. Also, a limit cycle may be important, and analysis or stabilization of a limit cycle is interesting to research. In order to tackle such problems, it is worth developing empirical nonlinear model reduction methods, which are also applicable around a non steady-state. Recently, a connection between POD and nonlinear controllability functions is established by~\cite{Kashima:16} in a stochastic setting. Empirical nonlinear observability Gramians have also been proposed~\cite{KI:09,PM:15}. Nevertheless, neither of these two methods deals with both controllability and observability Gramians, and there is no direct connection between these two works.

In this paper, we propose an empirical balancing method for nonlinear systems with linear time-invariant input vector field components by utilizing its variational system. Since the variational system can be viewed as a linear time-varying system (LTV) along the trajectory of the nonlinear system, one can extend the concept of the controllability and observability Gramians of the LTV system~\cite{KS:19,VK:83}. We call them the differential reachability and observability Gramians, respectively. They depend on the state trajectory of the nonlinear system. In general, it is not easy to obtain them as functions of the trajectory. Nevertheless, we show that their values at each fixed trajectory can be computed from the impulse and initial state responses of the variational system along this fixed trajectory. These obtained trajectory-wise Gramians are constant matrices, and thus one can compute balanced coordinates and a reduced order model in a similar manner as in the LTI case. 

The proposed empirical balancing method requires the variational system model. For large-scale systems, computing it may be challenging. Therefore, we further develop approximation methods, which do not require the variational model. Our approach is based on the fact that the variational system is a state space representation of the Fr\'echet derivative of an operator defined by the nonlinear system, and we use its discretization approximation. For the observability Gramian, similar approximation methods are found in~\cite{KI:09,PM:15}. However, there has been no corresponding controllability Gramian, which has been a bottleneck for developing the corresponding balancing method.

Similar nonlinear balanced realizations are found in flow balancing~\cite{VG:00,VG:04,Verriest:08} and in differential balancing~\cite{KS:17} but they are not empirical methods and require solutions to nonlinear PDEs. Moreover,~\cite{KS:17} does not give the concept of a Gramian. A preliminary version of our work is found in~\cite{KS:IFAC17}. In this paper, we further develop the discretization approximation methods. Moreover, we newly propose another differential balancing method for a class of nonlinear systems, which only requires the impulse responses of the variational system.

The remainder of this paper is organized as follows. In Section~2, we provide comprehensive back ground of linear balanced truncation in order to help understanding the whole picture of this paper. In Section~3,
we define the differential reachability and observability Gramians and then a differentially balanced realization along a trajectory of the system. In Section~4, we show that the value of the differential reachability/observability Gramian can be computed by using the impulse/initial state responses of the variational system. Then, we develop approximation methods, which only require empirical data of the original nonlinear system. In Section~5, we study positive definiteness of the differential reachability and observability Gramians related with nonlinear local strong accessibility and local observability. Next, we propose another differential balancing method, which is further computationally oriented. In Section~6, an example demonstrates our method for an RL network. Finally in Section~6, we conclude the paper by summarizing our results.

\section{Review of Linear Empirical Balancing}
In this section, we summarize the results for balanced truncation of linear time-invariant~(LTI) systems (for more details, see, e.g.~\cite{Antoulas:05,WP:02}) in order to help understanding the whole picture of this paper. 

Consider the following SISO LTI system:
\begin{align*}
\left\{\begin{array}{l}
\dot x (t) = A x(t) + B u(t),\\
y(t) = C x(t),
\end{array}\right.
\end{align*}
where $x(t)\in\bR^n$ and $u(t),y(t)\in\bR$; $A \in \bR^{n\times n}$, $B \in \bR^n$, and $C^\top \in \bR^n$. Its general solution is
\begin{align}
x (t) = e^{A(t-t_0)} x(t_0) + \int_{t_0}^t B e^{A(t-\tau)} u(\tau) d\tau. \label{sys_sol}
\end{align}
Based on the general solution, the controllability and observability Gramians are defined as
\begin{align}
&G_c(t_0,t_f) := \int_{t_0}^{t_f} e^{A(t-t_0)} B B^{\rm T} e^{A^{\rm T}(t-t_0)} dt,\label{CGram}\\
&G_o(t_0,t_f) := \int_{t_0}^{t_f} e^{A^{\rm T}(t-t_0)} C^{\rm T} C e^{A(t-t_0)} dt.\label{OGram}
\end{align}
They are positive definite for finite interval~$[t_0,t_f]$ for~$t_f>t_0$ if and only if the system is controllable and observable.

Let assume that the system is exponentially stable. When~$t_0=0$ and $t_f \to \infty$, it is known that the eigenvalues of the product~$G_o(0,\infty)G_c(0,\infty)$ correspond to the Hankel singular values of the linear system. Furthermore, there is a change of coordinates~$z=Tx$ such that
\begin{align*}
T G_c(0,\infty) T^{\rm T} &= T^{-\rm T} G_o (0,\infty) T^{-1}\\
& ={\rm diag}\{\sigma_1,\dots,\sigma_n\}, \ \sigma_i \ge \sigma_{i+1}.
\end{align*}
In this coordinate, $z_i$ are sorted in descending order corresponding to the Hankel singular values $\sigma_i$. That is,~$z_i$ is more important to capture the input-output behavior than $z_{i+1}$ if $\sigma_i>\sigma_{i+1}$. In balanced truncation, to approximate the input-output behavior by a reduced order model, the state variables corresponding to small Hankel singular values are truncated.

It is possible to compute the controllability/observability Gramian based on the impulse/initial state responses. From~\eqref{sys_sol}, the impulse response of the linear system is~$x_{\rm Imp}(t) = B e^{A(t-t_0)}$. From~\eqref{CGram}, one notices that
\begin{align*}
G_c(t_0,t_f) = \int_{t_0}^{t_f} x_{\rm Imp}(t) x_{\rm Imp}^{\rm T}(t) dt.
\end{align*}
Next, let $e^n_i \in \bR^n$ denote the standard basis, i.e., whose $i$th element is $1$, and the other elements are zero, and let $y_{{\rm Is},i}(t)$ denote the corresponding output response. Then, we have
\begin{align*}
y_{\rm Is}(t):=[\begin{array}{ccc}
y_{{\rm Is},1}(t) &\cdots &y_{{\rm Is},n}(t)
\end{array}]
= C e^{A(t-t_0)}.
\end{align*}
Moreover, from~\eqref{OGram}, one notices that
\begin{align*}
G_c(t_0,t_f) = \int_{t_0}^{t_f} y_{\rm Is}^{\rm T}(t) y_{\rm Is}(t) dt.
\end{align*}
Therefore, balanced truncation can be achieved based on empirical data. In this paper, we consider to extend these results to nonlinear systems.

\section{Differential Balancing along a Trajectory}
We present an empirical balancing method for a nonlinear system with LTI input vector field components by using its variational system; the reason considering such a vector field is elaborated in Remark~\ref{conip:rem} below. The proposed empirical balancing method is based on two Gramians, which we call differential reachability and observability Gramians. They can be viewed as extensions of Gramians for linear time-varying (LTV) systems~\cite{KS:19,VK:83} because the variational system can be viewed as an LTV system along a trajectory of the nonlinear system.

\subsection{Preliminaries}
Consider the following nonlinear system with LTI input vector field components (i.e., the input vector fields are constants):
\begin{align*}
\Sigma:
\left\{\begin{array}{l}
\dot x (t) = f(x(t)) + B u(t),\\
y(t) = h(x(t)),
\end{array}\right.
\end{align*}
where $x(t)\in\bR^n$, $u(t)\in\bR^m$, and $y(t)\in\bR^p$; $f: \bR^n \to \bR^n$ and $h: \bR^n \to \bR^p$ are of class $C^2$, and $B\in\bR^{n\times m}$. Let $\varphi_{t-t_0}(x_0,u)$ denote the state trajectory $x(t)$ of the system $\Sigma$ starting from $x(t_0)=x_0\in\bR^n$ for each choice of $u\in L_2^m[t_0,\infty)$. Note that since~$f$ is of class~$C^2$, if $u$ is also of class~$C^2$, then the solution~$\varphi_{t-t_0}(x_0,u)$ is a class~$C^2$ function of~$(t,x_0)$ as long as it exists. Throughout the paper, we assume that~$(\varphi_{t-t_0}(x_0,u),u(t))$ are of class~$C^2$ in a finite time interval $[t_0,t_f]$.

In our method, we use the prolonged system~\cite{CVC:05} of the system $\Sigma$, which consists of the original system $\Sigma$ and its variational system $d\Sigma$ along $x(t)=\varphi_{t-t_0}(x_0,u)$,
\begin{align*}
	d\Sigma:
	\left\{\begin{array}{l}
		\displaystyle\delta \dot x(t):=\frac{d \delta x(t)}{dt} = \frac{\partial f(\varphi_{t-t_0})}{\partial x} \delta x(t) + B \delta u(t),\\[2mm]
		\displaystyle \delta y(t) = \frac{\partial h(\varphi_{t-t_0})}{\partial x} \delta x(t),
	\end{array}\right.    
\end{align*}
where $\delta x(t)\in\bR^n$, $\delta u(t)\in\bR^m$ and $\delta y(t)\in\bR^p$. In the time interval $[t_0,t_f]$, the solution $\delta x(t)$ exists for any bounded input $\delta u(t)$ because the variational system $d\Sigma$ is an LTV system along $\varphi_{t-t_0}(x_0,u)$.

Since the variational system is an LTV system, it is possible to extend the aforementioned linear empirical balancing method to a nonlinear system via the variational system. To this end, we compute the solution $\delta x(t)$ of $d\Sigma$. It follows from the chain rule that
\begin{align}
\frac{d}{dt}\frac{\partial \varphi_{t-\tau}(x_{\tau},u)}{\partial x_{\tau}}
&= \frac{\partial}{\partial x_{\tau}}\frac{d \varphi_{t-\tau}(x_{\tau},u)}{d t}\nonumber\\
&= \frac{\partial f(\varphi_{t-\tau}(x_{\tau},u))}{\partial x_{\tau}}\nonumber\\
&=\frac{\partial f(\varphi_{t-\tau}(x_{\tau},u))}{\partial \varphi_{t-\tau}} \frac{\partial \varphi_{t-\tau}(x_{\tau},u)}{\partial x_{\tau}} \label{transition}
\end{align} 
That is, $\partial \varphi_{t-\tau}(x_{\tau},u)/\partial x_{\tau}$ is the transition matrix of $\partial f(\varphi_{t-\tau})/\partial x$ as an LTV system. From the general solution of an LTV system, the solution $\delta x(t)$ to the variational system $d\Sigma$ starting from $\delta x(t_0)=\delta x_0$ with input $\delta u(t)$ along the trajectory $\varphi_{t-t_0}(x_0,u)$ is obtained as
\begin{align}
\delta x(t) = \frac{\partial \varphi_{t-t_0}(x_0,u)}{\partial x} \delta x_0 + \int_{t_0}^t  \frac{\partial \varphi_{t-\tau}(x(\tau),u)}{\partial x} B \delta u(\tau) d\tau. \label{sol}
\end{align}
For the analysis, furthermore, we use a corresponding output when $\delta u \equiv 0$, namely
\begin{align}
\delta y(t) = &\frac{\partial h(\varphi_{t-t_0}(x_0,u))}{\partial x}  \frac{\partial \varphi_{t-t_0}(x_0,u)}{\partial x} \delta x_0. 
\label{osol}
\end{align}
%%%%%%%%%%%%%%%%%%%%%%%%%%%

\subsection{Differential Balanced Realization}
Inspired by results for LTI or LTV systems~\cite{KS:19,VK:83}, we define the differential reachability and observability Gramians from the variational systems as follows.
\begin{secdefn}
For given $x_0\in\bR^n$ and $u\in L_2^m[t_0,t_f]$, the differential reachability Gramian is defined as 
	\begin{align}
	G_{\cR}(t_0,t_f,x_0,u):= \int_{t_0}^{t_f} \frac{\partial \varphi_{t-t_0}}{\partial x} B B^{\rm T} \frac{\partial^{\rm T} \varphi_{t-t_0}}{\partial x} dt, \label{drG}
	\end{align}
where the arguments of~$\varphi_{t-t_0}$ are~$(x_0,u)$.
\end{secdefn}
\begin{secdefn}
For given $x_0\in\bR^n$ and $u\in L_2^m[t_0,t_f]$, the differential observability Gramian is defined as 
\begin{align}
&G_{\cO}(t_0,t_f,x_0,u) \nonumber\\
&:= \int_{t_0}^{t_f} \frac{\partial^{\rm T} \varphi_{t-t_0}}{\partial x} \frac{\partial^{\rm T} h(\varphi_{t-t_0})}{\partial \varphi_{t-t_0}}  \frac{\partial h(\varphi_{t-t_0})}{\partial  \varphi_{t-t_0}} \frac{\partial \varphi_{t-t_0}}{\partial x} dt, \label{doG}
\end{align}
where the arguments of~$\varphi_{t-t_0}$ are~$(x_0,u)$.
\end{secdefn}

Note that in the LTI case, they respectively reduce to the controllability Gramian~\eqref{CGram} and observability Gramian~\eqref{OGram}. These differential Gramians exist in $[t_0,t_f]$, $t_f>t_0$ from the assumption that the solution~$\varphi_{t-t_0}(x_0,u)$ exists and is of class $C^2$ in $[t_0,t_f]$.

\begin{secrem}
Our differential Gramians can be viewed as extensions of Gramians for LTV systems \cite{KS:19,VK:83}. By substituting $t = t_f + t_0 - \tau$ into~\eqref{drG}, we have
\begin{align*}
G_{\cR}(t_0,t_f,x_0,u)= \int_{t_0}^{t_f} &\frac{\partial \varphi_{t_f-\tau}}{\partial x} B B^{\rm T}  \frac{\partial^{\rm T} \varphi_{t_f-\tau}}{\partial x} d\tau,
\end{align*}
where $\varphi_{t_f-\tau}(x_f,{\mathcal F}_-(u))$ is the backward trajectory of the system $\Sigma$ starting from $x(t_f)=x_f$ with the input ${\mathcal F}_-(u)=u(t_f+t_0-\tau) \in L_2^m[t_0,t_f]$. This is an extension of the reachability Gramian for an LTV system in \cite{VK:83} to nonlinear prolonged systems. Similarly, the differential observability Gramian is an extended concept of the observability Gramian for LTV systems. \red
\end{secrem}

In a similar manner as a standard procedure, one can define a balanced realization between the differential reachability and observability Gramians. Since these differential Gramians are defined as functions of~$\varphi(x_0,u)$, we define our balanced realization trajectory-wise as follows.

\begin{secdefn}\label{DBR:def}
Let the differential reachability Gramian $G_{\cR}(t_0,t_f,x_0,u)\in\bR^{n \times n}$ and differential observability Gramian $G_{\cO}(t_0,t_f,x_0,u)\in\bR^{n \times n}$ at fixed $\varphi_{t-t_0}(x_0,u)$ be positive definite. A realization of the system $\Sigma$ is said to be a differentially balanced realization along $\varphi_{t-t_0}(x_0,u)$ if there exists a constant diagonal matrix
\begin{align*}
\Lambda ={\rm diag}\{\sigma_1,\dots,\sigma_n\}, \ \sigma_1 \ge \cdots \ge \sigma_n>0
\end{align*}
such that $G_{\cR}(t_0,t_f,x,u)=G_{\cO}(t_0,t_f,x,u)=\Lambda$.
\end{secdefn}

It is possible to show that there always exists a differentially balanced realization along $\varphi_{t-t_0}(x_0,u)$ if the differential Gramians are positive definite. Positive definiteness of them will be discussed in Section~\ref{PDG:s} related with local strong accessibility and local observability of the nonlinear system $\Sigma$.
\begin{secthm}\label{EDB:thm}
Suppose that differential Gramians $G_{\cR}(t_0,t_f,x_0,u)$ and $G_{\cO}(t_0,t_f,x_0,u)$ at fixed $\varphi_{t-t_0}(x_0,u)$ are positive definite. Then, there exists a non-singular matrix $T_{\varphi}\in\bR^{n \times n}$ which achieves 
\begin{eqnarray*}
T_{\varphi} G_{\cR}(t_0,t_f,x_0,u) T_{\varphi}^{\rm T} &=& T_{\varphi}^{-\rm T} G_{\cO}(t_0,t_f,x_0,u) T_{\varphi}^{-1} = \Lambda.
\end{eqnarray*}
That is, a differentially balanced realization along $\varphi_{t-t_0}(x_0,u)$ is obtained after a coordinate transformation $z=T_{\varphi}x$. \red
\end{secthm}

Since $G_{\cR}(t_0,t_f,x_0,u)$ and $G_{\cO}(t_0,t_f,x_0,u)$ are constant matrices, it is possible to prove Theorem~\ref{EDB:thm} in a similar manner as for the LTI case~\cite{Antoulas:05}. As in the LTI case, one can compute a reduced order model by truncating the state variables~$z_k,z_{k+1},\dots, z_n$ corresponding to small~$\sigma_k,\sigma_{k+1},\dots, \sigma_n$. Clearly, a reduced order model changes for a different trajectory and time interval. 

\section{Empirical Methods}
\subsection{Empirical Differential Gramians}
In the previous section, we defined a differentially balanced realization along a fixed trajectory $\varphi_{t-t_0}(x_0,u)$. For computing the differential Gramians as functions of $\varphi_{t-t_0}(x_0,u)$, or equivalently $(x_0,u)$, one needs to solve nonlinear partial differential equations (nPDEs) as for similar nonlinear balancing methods~\cite{KS:17,VG:00,VG:04,Verriest:08} in general. Hereafter, we focus on computing the values of the differential Gramians trajectory-wise.

First, we show that the differential reachability Gramian $G_{\cR}(t_0,t_f, \allowbreak x_0,u)$ along a fixed trajectory $\varphi_{t-t_0}(x_0,u)$ can be computed by using an impulse response of the variational system $d\Sigma$. Let~$\delta_D (\cdot)$ be Dirac's delta function, and let~$\delta x_{\rm Imp,i}(t)$ be the impulse response of the variational system~$d\Sigma$ along the trajectory~$\varphi_{t-t_0}(x_0,u)$ with $\delta u(t)=e^m_i \delta_D(t - t_0)$, where~$e^m_i \in \bR^m$ is the standard basis. Then, by substituting~$\delta x_0=0$ and $u(t)=e^m_i \delta_D(t - t_0)$ into~\eqref{sol}, we have 
\begin{align}
\delta x_{{\rm Imp},i}(t) = \frac{\partial \varphi_{t-t_0}(x_0,u)}{\partial x} B_i, \label{impbck}
\end{align}
where~$B_i$ is the~$i$th column vector of~$B$. From (\ref{drG}), we obtain
\begin{align}
&G_{\cR}(t_0,t_f,x_0,u) = \int_{t_0}^{t_f} \delta x_{\rm Imp}(t) \delta x_{\rm Imp}^{\rm T}(t) dt,\\
&\hspace{5mm} \delta x_{\rm Imp}(t) :=[\begin{array}{ccc}\delta x_{{\rm Imp},1}(t)& \cdots & \delta x_{{\rm Imp},m}(t) \end{array}].\nonumber
\end{align}
Therefore, for each $x_0\in\bR^n$ and $u\in L_2^m[t_0,t_f]$, the value of the differential reachability Gramian $G_{\cR}(t_0,t_f,x_0,u)$ is obtained by using the impulse response of $d\Sigma$.

\begin{secrem}\label{conip:rem}
The equality (\ref{impbck}) does not hold if $B$ is not constant. Indeed, for the system~$\dot x = f(x,u)$ and its trajectory~$\psi_{t-t_0}(x_0,u)$, the differential reachability Gramian is
	\begin{align*}
	&\bar G_{\cR}(t_0,t_f,x_0,u) \\
	&= \int_{t_0}^{t_f} \frac{\partial \psi_{t-t_0}}{\partial x} \frac{\partial f(\psi_{t-t_0},u)}{\partial u} \frac{\partial^{\rm T} f(\psi_{t-t_0},u)}{\partial u} \frac{\partial^{\rm T} \psi_{t-t_0}}{\partial x} dt.
	\end{align*}
However, the impulse response of the corresponding variational system is
\begin{align*}
&\delta \bar x_{\rm Imp}(t) \\
&=
\int_{t_0}^t \frac{\partial \psi_{t-\tau}(x(\tau ),u)}{\partial x} \frac{\partial f(\psi_{\tau-t_0}(x_0,u),u)}{\partial u} \delta_D(\tau - t_0) d\tau\\
&=
\frac{\partial \psi_{t-t_0}}{\partial x} \frac{\partial f(x_0,u(t_0))}{\partial u}.
\end{align*}
The reachability Gramian and impulse response do not coincide with each other for non-constant~$B$. \red
\end{secrem}

Next, we show that the differential observability Gramian $G_{\cO}(t_0,t_f,x_0,u)$ along a fixed trajectory $\varphi_{t-t_0}(x_0,u)$ can be computed by using initial state responses. By substituting~$\delta x_0=e^n_i$ and $\delta u = 0$ into~\eqref{osol}, one obtains the initial output response of $d\Sigma$ along $\varphi_{t-t_0}(x_0,u)$ as
\begin{align}
\delta  y_{{\rm Is},i}(t) = \frac{\partial h(\varphi_{t-t_0}(x_0,u))}{\partial x} \frac{\partial \varphi_{t-t_0}(x_0,u)}{\partial x} e^n_i, \label{initial}
\end{align}
From (\ref{doG}), we obtain
\begin{align*}
&G_{\cO}(t_0,t_f,x_0,u)= \int_{t_0}^{t_f}
\delta y_{\rm Is}^{\rm T}(t) \delta y_{\rm Is}(t) dt,\\
&\hspace{5mm}\delta y_{\rm Is}(t):=[\begin{array}{ccc}
\delta  y_{{\rm Is},1}(t) &\cdots &\delta  y_{{\rm Is},n}(t)
\end{array}].
\end{align*}
Thus, for each $x_0\in\bR^n$ and $u\in L_2^m[t_0,t_f]$, the value of the differential observability Gramian $G_{\cO}(t_0,t_f,x_0,u)$ is obtained by using the initial state response of $d\Sigma$.

In summary, the value of the differential reachability/observability Gramian for given~$x_0$ and~$u$ is obtained by computing impulse/initial state responses of a variational system $d\Sigma$ along the trajectory~$\varphi_{t-t_0}(x_0,u)$. Therefore, trajectory-wise differential balanced truncation is doable based on empirical data. 

\subsection{Approximation of the Fr\'echet Derivative}\label{DAFD:ss}
The empirical approach in the previous subsection requires the variational system model in addition to the original system model. If the original nonlinear systems are large-scale, computing the variational system model may need an effort. Therefore, we present approximation methods not requiring the variational system model.

In order to be self-contained, we first introduce the Fr\'echet derivative of a nonlinear operator. Consider a nonlinear operator $\Sigma (x_0,u): \bR^n \times L_2[t_0,t_f] \ni (x_0,u) \mapsto (x_f,y)\in \bR^n\times L_2[t_0,t_f]$ defined by the system $\Sigma$. A linear operator $d\Sigma_{(x_0,u)}(\delta x_0,\delta u)$ is said to be the Fr\'echet derivative if for each $x_0\in\bR^n$ and $u\in L_2[t_0,t_f]$, the following limit exists
\begin{align*}
&d\Sigma_{(x_0,u)} (\delta x_0,\delta u) \\
&:=  \lim_{s \to 0} \frac{\Sigma (x_0+s \delta x_0,u+s \delta u)- \Sigma (x_0,u)}{s}
\end{align*}
for all $\delta x_0\in\bR^n$ and $\delta u\in L_2[t_0,t_f]$. From its definition, the Fr\'echet derivative of nonlinear operator $\Sigma (x_0,u)$ is given by the variational system $d\Sigma$. Therefore, by using approximation of the Fr\'echet derivative, one can approximately compute the impulse or initial state responses of the variational systems. A simple approximation is
\begin{align*}
& d\Sigma_{(x_0,u)} (\delta x_0,\delta u)\\
& \approx d\Sigma_{(x_0,u)}^{\rm app} (\delta x_0,\delta u) :=\frac{\Sigma (x_0+s \delta x_0,u+s \delta u)- \Sigma (x_0,u)}{s}.
\end{align*}
Since the nonlinear operator $\Sigma(x_0,u)$ is given by the system $\Sigma$, a state space representation of the discretized approximation $d\Sigma_{(x_0,u)}^{\rm app} (\delta x_0,\delta u)$ is obtained as follows.
\begin{align*}
	d&\Sigma_{(x_0,u)}^{\rm app}(\delta x_0,\delta u):\\
	&\bR^n \times L_2^m[t_0,t_f] \times \bR^n \times L_2^m[t_0,t_f] \to \bR^n \times L_2^p[t_0,t_f],\\
	&(x_0,u,\delta x_0,\delta u) \mapsto (x_{vf},y_v),\\
	&\left\{\begin{array}{l}
		\begin{array}{l}
		\dot x^1(t)= f(x^1(t)) + B u^1(t), \\ \hspace{15mm}x^1(t_0)=x_0, \ u^1(\cdot) = u(\cdot)\\
		\dot x^2(t)= f(x^2(t)) + B u^2(t), \\ \hspace{15mm}x^2(t_0)=x_0+ s \delta x_0, \ u^2(\cdot) = u(\cdot)+s \delta u(\cdot)\\
		\displaystyle x_{vf}=\frac{x^2(t_f)-x^1(t_f)}{s},\ y_v(t)=\frac{h(x^2(t)) - h(x^1(t))}{s}.
		\end{array}
	\end{array}\right.
\end{align*}
Therefore,~$\delta x(t)$ and~$\delta y(t)$ can be approximately computed as~$\delta x(t) \simeq (x^2(t)-x^1(t))/s$ and~$\delta y(t) \simeq y_v(t)$, where~$\delta x_0$ and $\delta u$ coincide with the differences of pairs of the initial states~$(x^2(t_0)- x^1(t_0))/s$ and inputs~$(u^2- u^1)/s$, respectively.

From the above discussion, an approximation of the impulse response (\ref{impbck}) is obtained as
\begin{align*}
\delta x_{\rm Imp,i}(t) \approx \frac{x^2(t)-x^1(t)}{s}, \ \delta x_0 = 0, \delta u = e^m_i \delta_D (t-t_0),\\
i=1,2,\dots,m.
\end{align*}
Similar to the reachability Gramian, we need $m+1$ trajectories of the original nonlinear system in this computation by changing~$x^2(t)$ depending on the choice of~$\delta u$.

Next, an approximation of the initial state response (\ref{initial}) is
\begin{align*}
\delta y_{{\rm Is},i}(t) \approx y_v(t), \ \delta x_0 = e^n_i, \delta u =0, \ i=1,\dots,n.
\end{align*}
In this computation, we need $n+1$ trajectories of the original nonlinear system.

In summary, the differential reachability and observability Gramians can be approximately computed by using $n + m + 1$ trajectories of the original nonlinear system, where~$x^1(t)$ is same for the approximations of both differential reachability and observability Gramians. Note that even if one does not have an exact model of a real-life system, one only needs the impulse and initial state responses. Therefore, it may be possible to compute an approximation of a differentially balanced realization along $\varphi_{t-t_0}(x_0,u)$ by empirical data.

By applying our empirical methods, a change of coordinates~$z=T_{\varphi}x$ for balanced realization is obtained, and $T_{\varphi}$ depends on $\varphi_{t-t_0}(x_0,u)$. Still it is challenging to construct a reduced order model, which gives a good approximation for the whole trajectories because this essentially requires solving nPDEs. A potential solution to this problem is to apply deep learning techniques. After computing $T_{\varphi}$ for different choices of $\varphi_{t-t_0}(x_0,u)$, a function fitting method gives a global nonlinear change of coordinates for model reduction. An obtained reduced order model gives a good approximation at least around $\varphi_{t-t_0}(x_0,u)$ used for the computation of $T_{\varphi}$. We can
take arbitrary many trajectories, thus resulting in an approximate global method for model reduction. Another potential solution is to employ a basic idea of proper orthogonal decomposition. First, we compute the summation of differential reachability/observability Gramian, e.g. $(1/r)\sum_{i=1}^r G_{\cR}(t_0,t_f,x_i,u_i)$ for different choices of $\varphi_{t-t_0}(x_i,u_i)$, $i=1,\dots,r$. Then, we construct a linear change of coordinates which simultaneously diagonalize them and use this for truncation.

\subsection{Literature Review}
In literature, there are similar nonlinear balancing methods. We compare our methods with them.

First, another type of differential balancing method~\cite{KS:17} employs the following differential controllability and observability functions $L_{\cC}$ and $ L_{\cO}$.
\begin{align}
        L_{\cC}(x_0,u,\delta x_0):=\inf_{\delta u\in L_2^m(-\infty,t_0]}\frac{1}{2}\int_{-\infty}^{t_0} \|\delta u(t)\|^2 dt,\label{dcf}
\end{align}
where $x(t_0)=x_0\in\bR^n$, $u\in L_2^m(-\infty,t_0]$, $\delta x(t_0)=\delta x_0\in\bR^n$ and $\delta x(-\infty)=0$. 
\begin{align*}
        L_{\cO}(x_0,\delta x_0):=\frac{1}{2}\int^{\infty}_{t_0} \|\delta y(t)\|^2 dt,
\end{align*}
where $x(t_0)=x_0\in\bR^n$, $\delta x(t_0)=\delta x_0\in\bR^n$, $\delta x(\infty)=0$,~$u(t) \equiv 0$, and~$\delta u(t) \equiv 0$. Note that the differential controllability function gives the minimum energy to transfer the state of the prolonged system from $\delta x(-\infty)=0$ to $\delta x(t_0)=\delta x_0$ for given $x(t_0)=x_0$ and $u$. Therefore, it depends on~$x_0$,~$u$, and~$\delta x_0$. A similar discussion holds for the differential observability function.

In fact, by using~\eqref{osol} and~\eqref{doG}, the differential observability function and our differential observability Gramian are directly related as
\begin{align*}
L_{\cO}(x_0,u,\delta x_0)=\lim_{t_f \to \infty} \frac{1}{2} \delta x_0^{\rm T} G_{\cO}(t_0,t_f,x_0,u) \delta x_0.
\end{align*}
However, the differential reachability Gramian in (\ref{drG}) and the differential controllability function in (\ref{dcf}) are different. This corresponds to the difference between reachability and controllability of LTV systems~\cite{VK:83}. Reachability is the property to transfer the state from zero to an arbitrary terminal state, and controllability is the property to transfer the state from an arbitrary initial state to zero; they are the same properties for LTI systems. Based on the controllability Gramian of LTV systems, we define the differential controllability Gramian as
\begin{align*}
G_{\cC} (t_0,x_0,u):=  \int^{t_0}_{-\infty}  \frac{\partial \varphi_{t_0-\tau}}{\partial x} B B^{\rm T} \frac{\partial^{\rm T} \varphi_{t_0-\tau}}{\partial x} d\tau,
\end{align*}
where the arguments of $\varphi_{t_0-\tau}$ are $(x(\tau),u)$.
If this differential controllability Gramian $G_{\cC}(t_0,x_0,u)$ exists and is positive definite, the differential controllability function $L_{\cC}(x_0,u,\delta x_0)$ can be described as
\begin{align*}
L_{\cC}(x_0,u,\delta x_0) = \frac{1}{2} \delta x_0^{\rm T} G_{\cC}^{-1}(t_0,x_0,u) \delta x_0.
\end{align*}
The differential controllability Gramian is defined by using a backward trajectory of the nonlinear system~$\Sigma$. In contrast, the differential reachability Gramian is based on a forward trajectory and is computationally oriented.

Relating with differential balancing, flow balancing is proposed by~\cite{VG:00,VG:04,Verriest:08}. For flow balancing, the reachability and observability Gramians are defined on different time intervals, and the input is fixed for any initial state. Thus, the Gramians for flow balancing are defined as functions of the initial states. In contrast, our differential Gramians also depend on the input trajectory in addition to the initial state. Moreover, in order to achieve flow balancing, solutions to PDEs are required. Our methods may be applicable to develop empirical methods for flow balancing, which is included in future work.

The papers~\cite{LMG:02,HE:02,MR:04,Himpe:18,HE:02-2,WP:02} extend linear empirical balancing methods to nonlinear systems by focusing on a steady-state and attract a lot of research interests as computationally tractable nonlinear model reduction methods. Except~\cite{MR:04}, these methods can be viewed as our method with an approximation of the Fr\'echet derivative at a steady-state, and~\cite{MR:04} gives an empirical method with differential controllability (not reachability) and observability Gramians. In other words, we provide interpretations of those methods in terms of the variational system and an approximation of the Fr\'echet derivative. For observability, similar Gramians as ours are found for non-control systems~\cite{KI:09} and control systems~\cite{PM:15}. However, those papers do not provide the explicit description of the Gramians by using the solution of the original system or an interpretation in terms of the Fr\'echet derivative and do not establish the corresponding controllability Gramian. 

This is the first paper to develop empirical nonlinear balancing methods, which releases the requirement of~$\varphi_{t-t_0}(x_0,u)$ being a steady state. This relaxation is beneficial to enlarge the class of applications such as analysis and stabilization of a limit cycle and reducing computational complexity of trajectory tracking controller design for an arbitrary trajectory. Furthermore, as in~\cite{LMG:02,HE:02,MR:04,Himpe:18,HE:02-2,WP:02}, one may use non-impulse or non-initial state responses for model reduction. These different choices of inputs or initial states enable us to deal with wider classes of model reduction problems such as in~\cite{HRA:11} although such methods may not be interpreted in terms of Gramians.

% % % % % % % % % % % % % % % % % % % % % % % % % % % % % % % % % % % % % % % %
\section{Further Analysis}
In this section, we give some remarks for differential balancing proposed in this paper. First, we study positive definiteness of differential reachability and observability Gramians in terms of nonlinear local strong accessibility and local observability when $u\equiv 0$. Next, we show that for a specific class of systems, one can achieve another empirical differential balancing only by using the impulse responses of the variational system.
 
\subsection{Positive Definiteness of Gramians along Autonomous System}\label{PDG:s}
The differentially balanced realization is defined for positive definite differential reachability and observability Gramians. In a specific case when $u\equiv 0$, the positive definiteness implies local accessibility and observability of the original nonlinear system $\Sigma$, and the converse is true for local observability; see e.g.~\cite{NS:90} for the definitions of local strong accessibility and local observability.
\begin{secthm}\label{lsa:thm}
Let $f(x)$ be of class $C^{\infty}$. Then, the system $\Sigma$ is locally strongly accessible if the differential reachability Gramian $G_{\cR}(t_1,t_2,x_0,u)$ is positive definite for any $x_0$ and $u\equiv 0$ for any subinterval $[t_1,t_2]\subset [t_0,t_f]$.
\end{secthm}
\begin{pf}
For the sake of the simplicity of the discussion, we consider the single input case. Throughout the proof, we use the fact that the variational system $d\Sigma$ along $\varphi_{t-t_0}(x_0,0)$ is an LTV system. The differential reachability Gramian $G_{\cR}(t_1,t_2,x_0,0)$ is nothing but the controllability Gramian~\cite{Weiss:65} in the sense of LTV systems. For LTV systems, it has been shown in~\cite{SM:67,Weiss:65} that the controllability Gramian is positive definite for any subinterval $[t_1,t_2]$ if and only if the LTV system satisfies the Kalman-like controllability rank condition~\cite{SM:67}; the discussion until here holds for the multiple-input system. In the single input case, the Kalman-like controllability rank condition~\cite{SM:67} becomes as follows for the variational system: there exists $r>0$ such that
\begin{align}
{\rm rank} \left[\begin{array}{cccc}
B, & ad_f B, & \cdots ,& ad_f^{r-1} B
\end{array}\right](x_0) = n, \label{acc}
\end{align}
where~$ad_f^0B= f$ and $ad_f^{r}B  :=[ad_f^{r-1}B \ B]:=(\partial B/\partial x)ad_f^{r-1}B -  (\partial ad_f^{r-1}B/\partial x)B$, $i=1,2,\dots$; see~\cite{NS:90}. In the multiple-input case, $[ad_f^{r-1}B_i,B_j]$, $i,j=1,\dots,m$ are also needed to be taken into account.  Condition~(\ref{acc}) is a sufficient condition for local strong accessibility with Lie algebra; see e.g.~\cite{NS:90}.  \red
\end{pf}

\begin{secrem}\label{consv:rem}
Theorem~\ref{lsa:thm} gives a sufficient condition for local strong accessibility in terms of the differential reachability Gramian. The gap between condition (\ref{acc}) and strong accessibility is that the other Lie brackets that appear in the local strong accessibility rank condition~\cite{NS:90}, e.g., $[B,[f,B]]$ are missing; for more details see e.g.~\cite{NS:90}. To cover such Lie brackets, one needs to compute multiple differential reachability Gramians by changing inputs. 

Now, we provide the sketch of the idea of using multiple differential reachability Gramians in the single input case. Consider two differential reachability Gramians $G_{\cR}(t_1,t_2,x_0,0)$ and $G_{\cR}(t_1,t_2,x_0,u_1)$, where~$u_1(t) =1$ for all~$t \ge t_0$. From the results on the controllability analysis of LTV systems~\cite{SM:67,Weiss:65}, one can confirm that if there exists a non-zero constant vector~$v \in \bR^n$ such that
\begin{align}
G_{\cR}(t_1,t_2,x_0,0) v =0, \ G_{\cR}(t_1,t_2,x_0,u_1) v =0
\end{align}
for any subinterval $[t_1,t_2]\subset [t_0,t_f]$ $(t_1 < t_2)$ then
\begin{align*}
&{\rm rank} \left[\begin{array}{ccccc}
B & ad_f B & \cdots & ad_f^{r-1} B & ad_f^{r-1}B -  [B,[f,B]]
\end{array}\right](x_0)\\
&={\rm rank} \left[\begin{array}{ccccc}
B & ad_f B & \cdots & ad_f^{r-1} B &  [B,[f,B]]
\end{array}\right](x_0) < n.
\end{align*}
To cover all Lie brackets that appear in the local strong accessibility rank condition, one needs to compute a large amount of differential reachability Gramians for different inputs. This could even be an infinite number.
\red
\end{secrem}

For observability, we have a similar result.
\begin{secthm}\label{obGram:thm}
Let $u\equiv 0$ and $\delta u \equiv 0$. Also let $f(x)$ and $h(x)$ be of class $C^{\infty}$. Suppose that the observability codistribution \cite{NS:90} of $\Sigma$ has a constant dimension. Then, the system $\Sigma$ is locally observable if and only if the differential observability Gramian $G_{\cO}(t_1,t_2,x_0,u)$ is positive definite for any $x_0$ and $u\equiv 0$ for any subintervals $[t_1,t_2]\subset [t_0,t_f]$.
\end{secthm}
\begin{pf}
When the observability codistribution has a constant dimension, a system $\Sigma$ with $u\equiv 0$ is locally observable if and only if the nonlinear observability rank condition holds for all initial states \cite{NS:90}. One can confirm that this nonlinear observability rank condition is nothing but the Kalman-like observability rank condition~ \cite{SM:67} for the variational system $d\Sigma$ along the trajectory $\varphi_{t-t_0}(x_0,0)$ as an LTV systems. That is, the system $\Sigma$ is locally observable if and only if its variational system is differentially observable~\cite{SM:67,Weiss:65} under the assumption for the rank of the observability codistribution. Furthermore, the LTV system is differentially observable if and only if its observability Gramian is positive definite for any subinterval $[t_1,t_2]$ \cite{Weiss:65}, which is nothing but $G_{\cO}(t_0,t_f,x_0,0)$. 
\qed
\end{pf}

\begin{secrem}
In Theorem~\ref{obGram:thm}, a necessary and sufficient condition for the local observability of an autonomous system is provided, which is a sufficient condition for the local observability of a control system. For conservativeness, a similar discussion as Remark~\ref{consv:rem} holds. That is, to make the method less conservative, one may need to compute a large amount of differential observability Gramian with different inputs~$u$.\red
\end{secrem}

The paper~\cite{PM:15} gives a sufficient condition for local observability for non-zero $u$. As well known for LTV systems, the differential reachability and observability Gramians along $\varphi_{t-t_0}(x_0,0)$ are positive definite if and only if the variational systems along $\varphi_{t-t_0}(x_0,0)$ is completely controllable and observable, respectively.  The above theorems connect complete controllability~\cite{SM:67} and observability~\cite{SM:67} of the variational system and nonlinear local strong accessibility and observability, respectively. In addition, the theorems provide an empirical method for checking nonlinear accessibility and observability because one can compute the differential reachability and observability Gramians along $\varphi_{t-t_0}(x_0,u)$ by using impulse and initial state responses of the variational system, respectively. 

\subsection{Another Differential Balancing Method for Variationally Symmetric Systems}\label{VSS:ss}
Balancing methods including ours require two Gramians in general. One is for controllability, and the other is for observability. However, for linear systems, there is a class of systems for which one Gramian is constructed from the other. Such systems are called symmetric~\cite{Antoulas:05,SA:02,KS:19}. Motivated by the results for symmetric systems, we develop another differential balancing method.

This symmetry concept is extended to nonlinear systems~\cite{IFS:11} and variational systems~\cite{KS:16}. We further extend the latter symmetry concept.
\begin{secdefn}
The system $\Sigma$ is said to be variationally symmetric if there exists a class $C^1$ and non-singular $S : \bR^n \to \bR^{n\times n}$ such that
\begin{align}
&\sum_{i=1}^n \frac{\partial S(x)}{\partial x_i} f_i(x) + S(x) \frac{\partial f(x)}{\partial x}  = \frac{\partial^{\rm T} f(x)}{\partial x} S(x), \label{fS}\\
&S(x) B = \frac{\partial^{\rm T} h(x)}{\partial x} \label{BhS}
\end{align}
hold.
\end{secdefn}

Even though~$B$ is constant, a variationally symmetric system can have a nonlinear output because~$S$ is a function. If~$S$ is constant, the output function should be linear for a system being variationally symmetric. Variational symmetry implies that after a change of coordinates $\delta z = S(x) \delta x$, the variational system becomes
\begin{align}
	\left\{\begin{array}{l}
		\displaystyle\delta \dot z(t)= \frac{\partial^{\rm T} f(x(t))}{\partial x} \delta x(t) + \frac{\partial^{\rm T} h(x)}{\partial x} \delta u(t),\\[2mm]
		\displaystyle \delta y(t) =  B^{\rm T} \delta z(t).
	\end{array}\right.    \label{var_sym}
\end{align}
In the LTI case, the system (\ref{var_sym}) is called the dual system of the original system $\Sigma$, and the variational symmetry property is called symmetry. Many physical systems such as mechanical systems and RL circuits have this property; see e.g.~\cite{Schaft:11}.

For a symmetric LTI system, the controllability (observability) Gramian of the dual system corresponds to the observability (controllability) Gramian of the dual system~\cite{Antoulas:05,SA:02,KS:19}. Motivated by this fact, we consider to achieve model reduction based on the differential reachability Gramians of the original system and the system (\ref{var_sym}). For a variationally symmetric system, these two differential reachability Gramians are connected to each other. A similar relation holds for the differential observability Gramians. We leave to the reader this up.
\begin{secthm}\label{VSS:thm}
For the variationally symmetric system $\Sigma$ with respect to $S$, the differential reachability Gramian of the system (\ref{var_sym}) satisfies
\begin{align*}
&G_{\cR}^*(t_0,t_f,x_0,u) \\
&= \int_{t_0}^{t_f} S^{\rm T}(\varphi_t) \frac{\partial \varphi_{t-t_0}}{\partial x} B B^{\rm T}   \frac{\partial^{\rm T} \varphi_{t-t_0}}{\partial x} S(\varphi_{t-t_0}) dt.
\end{align*}
for any $x_0\in\bR^n$ and $u\in L_2^m[t_0,t_f]$ if it exists, where the arguments of~$\varphi_{t-t_0}$ are~$(x_0,u)$.
\end{secthm}
\begin{pf}
Throughout this proof, we omit the arguments of~$f$, $h$ and $S$, which are all~$\varphi_{t-\tau}(x_{\tau},u)$.
By using (\ref{transition}) and (\ref{fS}), compute
\begin{align*}
\frac{d}{dt}&\left(S \frac{\partial \varphi_{t-\tau}}{\partial x_{\tau}}S^{-1}\right)\\
=&\left(\sum_{i=1}^n \frac{\partial S}{\partial x_i} f_i + S \frac{\partial f}{\partial x}\right) \frac{\varphi_{t-\tau}}{\partial x_{\tau}} S^{-1}=\frac{\partial^{\rm T} f}{\partial x} S \frac{\partial \varphi_{t-\tau}}{\partial x_{\tau}}S^{-1}.
\end{align*} 
Therefore, $S(\varphi_{t-\tau}) (\partial \varphi_{t-\tau}/\partial x_{\tau})S^{-1}$ is the transition matrix of (\ref{var_sym}). 

From (\ref{BhS}), the differential reachability Gramian of (\ref{var_sym}) satisfies
\begin{align*}
&G_{\cR}^*(t_0,t_f,x_0,u) \\
&= \int_{t_0}^{t_f} S \frac{\partial \varphi_{t-\tau}}{\partial x_{\tau}} S^{-1} \frac{\partial^{\rm T}h}{\partial \varphi_{t-t_0}}  \frac{\partial h}{\partial  \varphi_{t-t_0}} S^{-\rm T}\frac{\partial^{\rm T} \varphi_{t-\tau}}{\partial x_{\tau}} S^{\rm T}dt\\
&= \int_{t_0}^{t_f}  S \frac{\partial \varphi_{t-\tau}}{\partial x_{\tau}} B  B^{\rm T} \frac{\partial^{\rm T} \varphi_{t-\tau}}{\partial x_{\tau}} S^{\rm T} dt. 
\end{align*}
That completes the proof.
\qed
\end{pf}

In the linear case, the cross Gramian~\cite{KS:19,Himpe:18} is useful for analysis of symmetric systems. However, the concept of a cross Gramian is missing in the differential case. Similar to the value of the differential reachability Gramian of the original nonlinear system, that of the system (\ref{var_sym}) can be computed by using the impulse response of the variational systems $d\Sigma$ for the variationally symmetric system. The applicability of the theory developed in this section is illustrated by an example in Section~\ref{ex:sec}. In fact, the paper~\cite{WP:02} gives an efficient empirical method for computing the linear controllability Gramian, but not for observablity Gramian. This method can be extended to the computation of the differential reachability Gramians. Therefore, differential balancing based on the only differential reachability Gramian can be proceed computationally efficiently.

\begin{figure}[tb]
	\begin{center}
		\includegraphics[width=70mm]{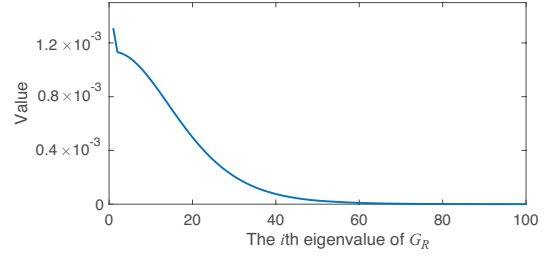}
		\caption{Eigenvalues of the differential reachability Gramian}
		\label{eig:fig}
	\end{center}
\end{figure}

%%%%%%%%%%%%%%%%%%%%%%%%%%%
\section{Example}\label{ex:sec}
As a simple example of variationally symmetric system, we consider the following nonlinear RL circuit with nonlinear resisters.
\begin{align*}
\left[\begin{array}{c}
\dot x_1 \\ \dot x_2 \\ \vdots \\ \dot x_{100}
\end{array}\right]
=& \left[\begin{array}{cccc}
-2 & 1 \\
1 & -2 & 1\\
& \ddots & \ddots & \ddots\\
& & 1 & -2 
\end{array}\right]
\left[\begin{array}{c}
	x_1 \\ x_2 \\ \vdots \\ x_{100}
\end{array}\right]\\
&-\left[\begin{array}{c}
x_1^2/2 + x_1^3/3 \\ x_2^2/2 + x_2^3/3 \\ \vdots \\ x_{100}^2/2 + x_{100}^3/3
\end{array}\right]+
\left[\begin{array}{c}
1 \\ 0 \\ \vdots  \\ 0
\end{array}\right] u,\\
y =& x_1.
\end{align*}
Note that due to the existence of $x_i^2$, we cannot use generalized differential balancing \cite{KS:15,KS:17} nor generalized incremental balancing \cite{BVSN:14}.

\begin{figure}[tb]
	\begin{center}
		\includegraphics[width=60mm]{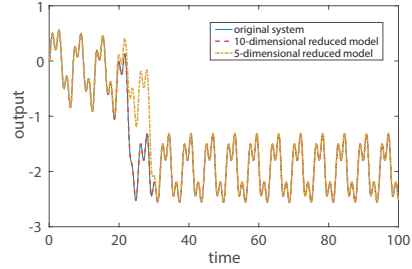}
		\caption{Output trajectories of $100$-dimensional original system and $5$ and $10$-dimensional reduced-order models}
		\label{RLsim:fig}
	\end{center}
\end{figure}

This system is variationally symmetric with respect to the identity matrix because $\partial f(x)/\partial x$ is symmetric. Thus, $G_{\cR} = G_{\cR}^*$ holds, i.e., we only have to compute the differential reachability Gramian $G_{\cR}$. We compute its value numerically based on the method in Section~\ref{DAFD:ss} with $s=0.01$ (we also try our numerical method with~$s=1$, and the obtained reduced order model is similar as for $s=0.01$ in this example). For this computation, we need snapshots of the trajectories of the system. Since the system is a single input system, we need two trajectories; one is~$x^1(t)$ around which a reduced oder model is constructed, and the other is~$x^2(t)$ needed for the computation of the impulse response. For instance, we choose~$x^1(t)$ as the trajectory starting from $x(0)=0$ with input $u = \sin(t) + \sin(3t)$. As a numerical computational method of snapshots, we use the forward Euler method with the step size~$\Delta t =0.01$. Then, the differential reachability Gramian~$G_{\cR}$ is computed numerically in the time interval~$[0,100]$. In the case when $G_{\cR} = G_{\cR}^*$, empirical differential balanced truncation is achieved by computing the eigenspace of~$G_{\cR}$ and truncating state variables corresponding to small eigenvalues. Figure~\ref{eig:fig} shows eigenvalues of~$G_{\cR}$.
Figure~\ref{RLsim:fig} shows the output trajectories of the original system, 10-and 5-dimensional reduced order models.  It can be observed that the trajectory of the 10th order model follows the trajectory of the original model really well. The 5th order model is still a good approximation except the time interval $[20,30]$ when the output trajectory changes significantly. Every process here is conduced by using Matlab 2019a on MacBook Pro (2.7 GHz Intel Core i7;16 GB 2133 MHz LPDDR3).
%---------------------------------------------------------
%---------------------------------------------------------
\section{Conclusion and Future Work}
\subsection{Conclusion}
In this paper, we have proposed a nonlinear empirical differential balancing method along a fixed state trajectory for nonlinear systems. Our method is based on the differential reachability and observability Gramians, which are functions of the state trajectory. The values of these Gramians at each trajectory is computable by using impulse and initial state responses of the variational system along the trajectory. We have also developed approximation methods for computing them, which only requires empirical data of the original nonlinear systems. 

\subsection{Possible Application}
In~\cite{HKE:02,CSB:16}, empirical balancing at a stead-state is used to reduce the computational complexity of nonlinear model predictive control (MPC) \cite{GP:11,CA:13}. Our proposed empirical differential balancing method along a fixed state trajectory can be used to reduce the computational complexity around non steady states. In MPC, we repeatedly solve the following nonlinear optimal control problem.
\begin{align}
J = \varphi (x(t+T),u(t+T)) + \int_t^{t+T} L(x(\tau),u(\tau)) d\tau.\label{opt}
\end{align}
If the optimal control input $u$ in the time interval $[t_0, t_0 + \Delta t]$ is obtained, one can compute the state trajectory of the controlled system in this time interval. Along this trajectory, it is possible to achieve the proposed empirical differential balanced truncation. Then, we have a reduced order model. To compute the optimal control input in the next time interval $[t_0 + \Delta t, t_0 + 2 \Delta t]$, one can use the reduced order model. For this reduced order model, one can compute the corresponding cost function to (\ref{opt}). By solving the reduced order optimal control problem, one has the optimal control input for the reduced order model, which is an approximation of the optimal control input for the original system in the time interval $[t_0 + \Delta t, t_0 + 2 \Delta t]$. Thus, one can use this input for controlling the original system and have the state trajectory of the controlled original system in the time interval $[t_0 + \Delta t, t_0 + 2 \Delta t]$. Then, one can again exploit our empirical model reduction method for obtaining a reduced order model. By repeating this procedure, one can compute an approximation of the optimal control input in each time interval recursively.

\section*{Acknowledgment}
This work of Y. Kawano was supported by JSPS KAKENHI Grant Number JP19K23517 and JST CREST Grant Number JPMJCR15K2, Japan.
% % % % % % % % % % % % % % % % % % % % % % % % %
\bibliographystyle{plain}
\bibliography{dbalref.bib}

\begin{thebibliography}{10}

\bibitem{Antoulas:05}
A.~C. Antoulas.
\newblock {\em Approximation of Large-Scale Dynamical Systems}.
\newblock SIAM, Philadelphia, 2005.

\bibitem{Astolfi:10}
A.~Astolfi.
\newblock Model reduction by moment matching for linear and nonlinear systems.
\newblock {\em IEEE Trans. Aut. Control}, 55(10):2321--2336, 2010.

\bibitem{BVSN:14}
B.~Besselink, N.~van~de Wouw, J.~M.~A. Scherpen, and H.~Nijmeijer.
\newblock Model reduction for nonlinear systems by incremental balanced
  truncation.
\newblock {\em IEEE Trans. Aut. Control}, 59(10):2739 -- 2753, 2014.

\bibitem{CA:13}
E.~F. Camacho and C.~B. Alba.
\newblock {\em Model Predictive Control}.
\newblock Springer Science \& Business Media, New York, 2013.

\bibitem{CSB:16}
R.~B. Choroszucha, J.~Sun, and K.~Butts.
\newblock Nonlinear model order reduction for predictive control of the diesel
  engine airpath.
\newblock {\em Proc. 2016 American Control Conference}, pages 5081--5086, 2016.

\bibitem{MR:04}
M.~Condon and R.~Ivanov.
\newblock Empirical balanced truncation of nonlinear systems.
\newblock {\em Journal of Nonlinear Science}, 14(5):405--414, 2004.

\bibitem{CVC:05}
J.~Cort{\'e}s, A.~J. van~der Schaft, and P.~E. Crouch.
\newblock Characterization of gradient control systems.
\newblock {\em SIAM J. Control and Optimization}, 44(4):1192--1214, 2005.

\bibitem{FS:05}
K.~Fujimoto and J.~M.~A. Scherpen.
\newblock Nonlinear input-normal realizations based on the differential
  eigenstructure of {H}ankel operators.
\newblock {\em IEEE Trans. Aut. Control}, 50(1):2--18, 2005.

\bibitem{FT:08}
K.~Fujimoto and D.~Tsubakino.
\newblock Computation of nonlinear balanced realization and model reduction
  based on {T}aylor series expansion.
\newblock {\em Sys. Cont. Lett.}, 57(4):283--289, 2008.

\bibitem{GP:11}
L.~Gr\"une and J.~Pannek.
\newblock {\em Nonlinear Model Predictive Control}.
\newblock Springer-Verlag, New York, 2011.

\bibitem{HE:02-2}
J.~Hahn and T.~F. Edgar.
\newblock Balancing approach to minimal realization and model reduction of
  stable nonlinear systems.
\newblock {\em Industrial \& Engineering Chemistry Research}, 41(9):2204--2212,
  2002.

\bibitem{HE:02}
J.~Hahn and T.~F. Edgar.
\newblock An improved method for nonlinear model reduction using balancing of
  empirical gramians.
\newblock {\em Computers \& Chemical Engineering}, 26(10):1379--1397, 2002.

\bibitem{HKE:02}
J.~Hahn, U.~Kruger, and T.~F. Edgar.
\newblock Application of model reduction for model predictive control.
\newblock {\em IFAC Proceedings Volumes}, 35(1):393--398, 2002.

\bibitem{HRA:11}
M.~Heinkenschloss, T.~Reis, and A.~C. Antoulas.
\newblock Balanced truncation model reduction for systems with inhomogeneous
  initial conditions.
\newblock {\em Automatica}, 47(3):559--564, 2011.

\bibitem{Himpe:18}
C.~Himpe.
\newblock emgr―the empirical gramian framework.
\newblock {\em Algorithms}, 11(7):91, 2018.

\bibitem{HLB:12}
Philip Holmes, John~L. Lumley, Gahl Berkooz, and Clarence~W. Rowley.
\newblock {\em Turbulence, Coherent Structures, Dynamical Systems and
  Symmetry}.
\newblock Cambridge University Press, 2012.

\bibitem{IA:16}
T.~C. Ionescu and A.~Astolfi.
\newblock Nonlinear moment matching-based model order reduction.
\newblock {\em IEEE Trans. Aut. Control}, 61(10):2837--2847, 2016.

\bibitem{IFS:11}
T.~C. Ionescu, K.~Fujimoto, and J.~M.A. Scherpen.
\newblock Singular value analysis of nonlinear symmetric systems.
\newblock {\em IEEE Trans. Aut. Control}, 56(9):2073--2086, 2011.

\bibitem{Kashima:16}
K.~Kashima.
\newblock Noise response data reveal novel controllability {G}ramian for
  nonlinear network dynamics.
\newblock {\em Scientific Reports}, 6(27300), 2016.

\bibitem{KBS:19}
Y.~Kawano, B.~Besselink, J.~M.~A. Scherpen, and M.~Cao.
\newblock Data-driven model reduction of monotone systems by nonlinear {DC}
  gains.
\newblock {\em IEEE Trans. Aut. Control}, 65(6), 2020.
\newblock (early access).

\bibitem{KS:15}
Y.~Kawano and J.~M.~A. Scherpen.
\newblock Model reduction by generalized differential balancing.
\newblock In M.~K. Camlibel, A.~A. Julius, R.~Pasumarthy, and J.~M.~A.
  Scherpen, editors, {\em Mathematical Control Theory I}. Springer-Verlag, pp.
  349-362, 2015.

\bibitem{KS:16}
Y.~Kawano and J.~M.~A. Scherpen.
\newblock Generalized differential balancing for variationally symmetric
  systems.
\newblock {\em IFAC-PapersOnLine}, 49(18):844--849, 2016.

\bibitem{KS:IFAC17}
Y.~Kawano and J.~M.~A. Scherpen.
\newblock Empirical differential balancing for nonlinear systems.
\newblock {\em IFAC-PapersOnLine}, 50(1):6326--6331, 2017.

\bibitem{KS:17}
Y.~Kawano and J.~M.~A. Scherpen.
\newblock Model reduction by differential balancing based on nonlinear {H}ankel
  operators.
\newblock {\em IEEE Trans. Aut. Control}, 62(7):3293--3308, 2017.

\bibitem{KS:19}
Y.~Kawano and J.~M.~A. Scherpen.
\newblock Balanced model reduction for linear time-varying symmetric systems.
\newblock {\em IEEE Trans. Aut. Control}, 64(7):3060--3067, 2019.

\bibitem{KI:09}
A.~J. Krener and K.~Ide.
\newblock Measures of unobservability.
\newblock {\em Proc. 48th IEEE Conference on Decision and Control and the 28th
  Chinese Control Conference}, pages 6401--6406, 2009.

\bibitem{LMG:02}
S.~Lall, J.~E. Marsden, and S.~Glava{\v{s}}ki.
\newblock A subspace approach to balanced truncation for model reduction of
  nonlinear control systems.
\newblock {\em International Journal of Robust and Nonlinear Control},
  12(6):519--535, 2002.

\bibitem{NK:00}
A.~J. Newman and P.~S. Krishnaprasad.
\newblock Computing balanced realizations for nonlinear systems.
\newblock Technical report, 2000.

\bibitem{NS:90}
H.~Nijmeijer and A.J. van~der Schaft.
\newblock {\em Nonlinear Dynamical Control Systems}.
\newblock Springer-Verlag, New York, 1990.

\bibitem{PM:15}
N.~D. Powel and K.~A. Morgansen.
\newblock Empirical observability {G}ramian rank condition for weak
  observability of nonlinear systems with control.
\newblock {\em Proc. 54th IEEE Conference on Decision and Control}, pages
  6342--6348, 2015.

\bibitem{SA:14}
M.~Sassano and A.~Astolfi.
\newblock Dynamic generalized controllability and observability functions with
  applications to model reduction and sensor deployment.
\newblock {\em Automatica}, 50(5):1349--1359, 2014.

\bibitem{SA:17}
G.~Scarciotti and A.~Astolfi.
\newblock Data-driven model reduction by moment matching for linear and
  nonlinear systems.
\newblock {\em Automatica}, 79:340--351, 2017.

\bibitem{Scherpen:93}
J.~M.~A. Scherpen.
\newblock Balancing for nonlinear systems.
\newblock {\em Sys. Cont. Lett.}, 21(2):143--153, 1993.

\bibitem{SM:67}
L.~M. Silverman and H.~E. Meadows.
\newblock Controllability and observability in time-variable linear systems.
\newblock {\em SIAM Journal on Control}, 5(1):64--73, 1967.

\bibitem{SA:02}
D.~C. Sorensen and A.~C. Antoulas.
\newblock The {S}ylvester equation and approximate balanced truncation.
\newblock {\em Linear Algebra and its Applications}, 351--352:671--700, 2002.

\bibitem{Schaft:11}
A.~J. van~der Schaft.
\newblock On the relation between port-{H}amiltonian and gradient systems.
\newblock {\em IFAC Proceedings Volumes}, 44(1):3321--3326, 2011.

\bibitem{Verriest:08}
E.~I. Verriest.
\newblock Time variant balancing and nonlinear balanced realizations.
\newblock {\em Model Order Reduction: Theory, Research Aspects and
  Applications}, pages 213--250, 2008.

\bibitem{VG:00}
E.~I. Verriest and W.~S. Gray.
\newblock Flow balancing nonlinear systems.
\newblock {\em Proc. 14th International Symposium on Mathematical Theory of
  Networks and Systems}, 2000.

\bibitem{VG:04}
E.~I. Verriest and W.~S. Gray.
\newblock Nonlinear balanced realizations.
\newblock {\em Proc. 43rd IEEE Conference on Decision and Control}, pages
  1164--1169, 2004.

\bibitem{VK:83}
E.~I. Verriest and T.~Kailath.
\newblock On generalized balanced realizations.
\newblock {\em IEEE Trans. Aut. Control}, 28(9):833--844, 1983.

\bibitem{Weiss:65}
L.~Weiss.
\newblock The concepts of differential controllability and differential
  observability.
\newblock {\em Journal of Mathematical Analysis and Applications}, 10(2), 1965.

\bibitem{WP:02}
K.~Willcox and J.~Peraire.
\newblock Balanced model reduction via the proper orthogonal decomposition.
\newblock {\em AIAA Journal}, 40(11):2323--2330, 2002.

\bibitem{ZDG:96}
K.~Zhou, J.~C. Doyle, and K.~Glover.
\newblock {\em Robust and Optimal Control}, volume~40.
\newblock Prentice Hall, New Jersey, 1996.

\end{thebibliography}
\end{document}